\documentclass[12pt,a4paper,notcite,notref]{article}

\usepackage{amsmath,amssymb,epsfig,epic}
\usepackage{color}
\usepackage{epstopdf}
\usepackage{graphicx}

\begin{document}
\title{Integrable lattice equations with vertex and bond variables}
\author{Jarmo Hietarinta$^{1,2}$\footnote{E-mail:
jarmo.hietarinta@utu.fi}~ and Claude Viallet$^{1}\footnote{E-mail:
claude.viallet@upmc.fr}~$  \\
{\small\it $^1$ LPTHE / CNRS / UPMC, 4 place Jussieu 75252 Paris CEDEX
  05, France }\\
{\small\it $^2$Department of Physics and Astronomy, University of
  Turku, FIN-20014 Turku, Finland\footnote{permanent address}}}

\maketitle

\begin{abstract}
  We present integrable lattice equations on a two dimensional square
  lattice with coupled vertex and bond variables. In some of the
  models the vertex dynamics is independent of the evolution of the
  bond variables, and one can write the equations as non-autonomous
  ``Yang-Baxter maps''. We also present a model in which the vertex
  and bond variables are fully coupled.  Integrability is tested with
  algebraic entropy as well as multidimensional consistency.
\end{abstract}

\section{Introduction}

Along with the classification results for integrable lattice maps on
quad-graphs with one component fields on vertices~\cite{ABS}, various
more general integrable lattice equations have been proposed recently. This
includes the ``Yang-Baxter maps'' (solutions to the functional
Yang-Baxter equations~\cite{Dr92}) with variables on bonds
\cite{Ve03,ABS-CAG,PaTo07}, and number of higher order as well as
multi-component cases \cite{JH-B,SpNiKa10,Ni11}.

We describe here integrable lattice models which have both vertex and
bond variables.\footnote{ The term ``lattice models with bond and
  vertex variables'' has already been used in the context of quantum
  spin models~\cite{JKMO}, but the setting is then quite different:
  the vertex and bond variables are independent variables and take on
  discrete values, the main data being the Boltzmann weight of the
  configuration, and one is the interested in the statistical
  mechanics of the system. Here, in contrast, the lattice equations
  define a 1+1 dimensional evolution of the variables.}  These models
were found in the analysis of a coarse graining process of known
integrable models on a square lattice \cite{HV11a}. We first describe
the basic procedure we used, and the way to test integrability via
algebraic entropy (section \ref{roadmap}). We then list various models
obtained in this way, giving explicitly the evolution they define and
their entropy. We also show that the evolution may be determined from
a set of algebraic relations which have special rationality properties
(section \ref{models}).  We finally examine the three dimensional
consistency of these models, and show that some of them can be
interpreted as solutions of non-autonomous functional Yang-Baxter
equations (section \ref{nonautoYB}).

\section{A road map} \label{roadmap}

\subsection{The setting} \label{setting}
The starting point is a regular planar square lattice. Its vertices
are labeled by two integers $(n,m)$. We introduce two kinds of
variables: the vertex variables, denoted $w_{n,m}$, and the bond
variables denoted $X_{n,m}$ for the horizontal bonds, and $Y_{n,m}$
for the vertical bonds. The bond variables are indexed by their base
vertex, which is to the left (respectively down) along the bond, as
shown in Figure \ref{Fdef}(a).  As usual we use the shorthand notation
where only shifts are indicated, e.g., $X_2=X_{n,m+1},
w_{12}=w_{n+1,m+1}$, etc...

The models are given by three relations between the eight variables
associated to each square cell, in such a way that they define an
evolution: from the three defining relations, it is possible to
calculate $X_2,Y_1,w_{12}$ (open circles in Figure \ref{Fdef}(a)) from
$w,X,Y,w_1,w_2$ (black disks).  To fully define an evolution, we have
to give a suitable set of initial data.  Such data may be given on a
diagonal staircase, after which the vertex and bond variables may be
evaluated on the entire lattice, using the local relations on each
cell.

We will be interested in models where the evolution is rational, i.e.,
$X_2,Y_1,w_{12}$ are rationally expressed in terms of $w,X,Y,w_1,w_2$.
For the models we will consider, one may actually permute the role of
the four corners, and in particular define a backward evolution
calculating $w,X,Y$ in terms of $Y_1,X_2,w_1,w_2,w_{12}$. This is a
form of rational invertibility, similar to birationality for maps. For
the models we describe, it is actually possible to define rational
evolution in all four directions on the lattice.

\subsection{A construction procedure}
There is one very simple way to generate such integrable maps from
known one-component lattice maps.

\begin{figure}[th!]
\begin{center}
\setlength{\unitlength}{0.0007in}
\begin{picture}(3482,2813)(0,-10)
\put(450,1883){\circle*{91}}
\put(1350,1883){\circle{91}}
\put(2250,1883){\circle{91}}
\put(2250,983){\circle{90}}
\put(450,983){\circle*{90}}
\put(2250,83){\circle*{90}}
\put(1350,83){\circle*{90}}
\put(450,83){\circle*{90}}
\put(0,1983){\makebox(0,0)[lb]{\small$w_{2}$}}
\put(2550,8){\makebox(0,0)[lb]{\small$w_{1}$}}
\put(2550,1983){\makebox(0,0)[lb]{\small$w_{12}$}}
\put(100,8){\makebox(0,0)[lb]{\small$w$}}
\put(100,908){\makebox(0,0)[lb]{\small$Y$}}
\put(-500,908){\makebox(0,0)[lb]{(a)}}
\put(2550,908){\makebox(0,0)[lb]{\small$Y_1$}}
\put(1200,-200){\makebox(0,0)[lb]{\small$X$}}
\put(1200,2083){\makebox(0,0)[lb]{\small$X_2$}}
\drawline(450,1883)(450,83)
\drawline(450,1883)(2250,1883)
\drawline(2250,1883)(2250,83)
\drawline(450,83)(2250,83)
\end{picture}
\begin{picture}(3482,2813)(-1000,-10)
\put(450,1883){\circle*{91}}
\put(1350,1883){\circle{91}}
\put(2250,1883){\circle{91}}
\put(2250,983){\circle{90}}
\put(450,983){\circle*{90}}
\put(2250,83){\circle*{90}}
\put(1350,83){\circle*{90}}
\put(450,83){\circle*{90}}
\put(0,1983){\makebox(0,0)[lb]{\small$x_{22}$}}
\put(2550,8){\makebox(0,0)[lb]{\small$x_{11}$}}
\put(2550,1983){\makebox(0,0)[lb]{\small$x_{1122}$}}
\put(100,8){\makebox(0,0)[lb]{\small$x$}}
\put(100,908){\makebox(0,0)[lb]{\small$x_2$}}
\put(-500,908){\makebox(0,0)[lb]{(b)}}
\put(2550,908){\makebox(0,0)[lb]{\small$x_{112}$}}
\put(1200,-200){\makebox(0,0)[lb]{\small$x_1$}}
\put(1200,2083){\makebox(0,0)[lb]{\small$x_{122}$}}
\put(1430,1050){\makebox(0,0)[lb]{\tiny$x_{12}$}}
\drawline(450,1883)(450,83)
\drawline(450,1883)(2250,1883)
\drawline(2250,1883)(2250,83)
\drawline(450,83)(2250,83)
\dashline{60.000}(1350,1883)(1350,83)
\dashline{60.000}(450,983)(2250,983)
\end{picture}
\end{center}
\caption{(a): the basic lattice square, and (b): its precursor, a $2\times2$
  sublattice. \label{Fdef}}
\end{figure}
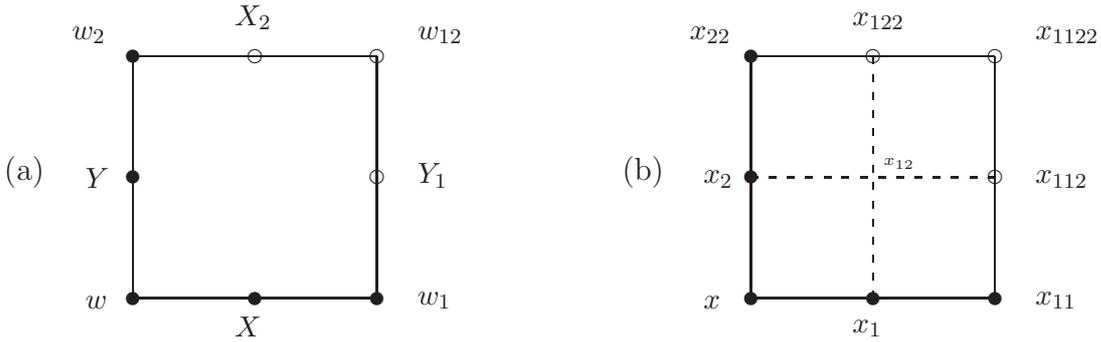

Suppose we are given an integrable quad map.  As shown in Figure
\ref{Fdef}(b), consider $\{x,x_{11},x_{22}, x_{1122}\}$ as vertex
variables of a coarser lattice (renaming them $\{w,w_1,w_2,w_{12}\}$,
the rule being $w_{n,m}=x_{2n,2m}$), and $\{x_2,x_{112},x_1,x_{122}\}$
as bond variables (and renaming them $\{Y,Y_1,X,X_2\}$, i.e.,
$X_{n,m}=x_{2n+1,2m},\,Y_{n,m}=x_{2n,2m+1}$). The quad map determines
$x_{12}$ in terms of $x,x_1,x_2$, and $x_{122}$ in terms of
$x_2,x_{22},x_{12}$, and so on, after which one may forget the central
$x_{12}$.  One signature of this construction is that $X_2$ does not
depend on $w_1$, and $Y_1$ does not depend on $w_2$.

The purpose of this paper is to present a number of models which
cannot be obtained in the simple way described above, and to analyze
their properties.  They will in particular satisfy
\begin{equation}\label{cond}
\frac{\partial  X_{n,m+1}}{\partial{w_{n+1,m}}} \cdot 
\frac{\partial  Y_{n+1,m}}{\partial{w_{n,m+1}}}
\neq 0, \,\forall n,m.
\end{equation}

In order to construct such maps we start from quad maps having a
$2\times 2$ Lax pair~\cite{FLax,BSLax}. We use this Lax pair to write
the zero curvature condition on the larger square of Figure
\ref{Fdef}(b). From the size of the Lax matrices, we see that this
zero curvature condition can give at most three scalar conditions. We
know that one rational solution can be obtained in the way described
above, we call this solution ``regular''. {\em What is remarkable is
  that for certain quad maps, there exists another rational solution},
which we called ``exotic''~\cite{HV11a}.

In other words, we use the fact that the zero curvature condition
written on a coarser lattice is ambiguous. In most cases there is only
one solution giving {\em rational} evolution, it is the regular
one. There are usually other solutions that are not rational, but we discard
them. Only in some exceptional cases do we find rational solutions
satisfying \eqref{cond}.\footnote{In some respects this is similar to
  using conservation laws to derive dynamics, which sometimes allows
  multiple solutions\cite{QCRdual}.}

\subsection{Testing integrability}
When an exotic solution exists, there is no guarantee a priori that it
leads to an integrable evolution, and we have to check its
integrability.

The setting is appropriate to use the vanishing algebraic entropy
criterion~\cite{AE1,AE2}.  Given initial data on a basic diagonal
staircase, one evaluates the $w$'s, $X$'s and $Y$'s away from the
initial diagonal in terms of these data.  The entropy can then be
extracted from the sequences $\{d_n\}$ of the successive degrees respectively for the  $w$'s, $X$'s and $Y$'s.  The entropy  is the limit
\begin{eqnarray*}
\eta = \lim_{n\rightarrow \infty }\frac{1}{n} \, \log ( d_n ).
\end{eqnarray*}
This limit always exists.  {\em The characteristic of integrability is
  the vanishing of $\eta$.} It is equivalent to polynomial growth
of the sequences $\{d_n\}$.  Non-vanishing of the entropy, that is to
say exponential growth of a sequence $\{d_n\}$, means
non-integrability.

\paragraph{Remark}: The simplest way to find the exact value of the
entropy is by fitting, if possible, the generating function $\zeta(s)=
\sum_n d_n s^n$ of the sequence of degrees by a rational fraction. The
entropy is then determined by the position of the poles of that
fraction~\cite{AE1,AE2}.

\subsection{Simplified defining equations}

We get the exotic solutions from a set of three algebraic conditions
equivalent to a zero curvature condition, having more than one
solution.  It is possible, for all the models we describe below, to
write three simpler equations, which have, {\em as a unique solution},
the exotic model.

These equations define a rational variety of dimension $5$ in an $8$
dimensional space. They have special (multi)-rationality properties:
for any choice of a corner in Figure \ref{Fdef}(a), the three
corresponding variables (one vertex variable and the two adjacent bond
variables) can be expressed rationally in terms of the five
others. This special rationality property allows one to define rational
evolutions in all four directions of the square lattice. This is a
generalized form of the notion of quadrirationality introduced
in~\cite{ABS-CAG}.

Equivalently, choosing two adjacent bonds, the relations define $2
\mapsto 2$ birational maps between the variables attached to these
bonds and the two remaining ones.  These maps and their inverses
actually happen to have the same form, but they are not involutions a
priori.

\section{The models} \label{models}

\subsection{dpKdV  (H1)\label{S3.1}}
The simplest case is obtained from the lattice potential KdV (H1
in ~\cite{ABS}). Computing the zero curvature condition on the coarser
lattice we found two rational solutions~\cite{HV11a}. We give here for
reference the regular one, which does not fulfill condition
\eqref{cond}:
\begin{subequations}\label{H1reg}
\begin{eqnarray}
X_2&=&Y+\frac{(q-p)(Y-X)}{(Y-X)(w-w_2)-(q-p)},\\
Y_1&=&X+\frac{(p-q)(X-Y)}{(X-Y)(w-w_1)-(p-q)},\\
w_{12}&=&w+(p-q)\frac{(p-q)(w_1+w_2-2w)+2(w-w_1)(w-w_2)(X-Y)}
{(p-q)^2-(w-w_1)(w-w_2)(X-Y)^2}.
\end{eqnarray}
\end{subequations}
 The  exotic solution is:
\begin{subequations}\label{H1ex-ep0}
\begin{eqnarray}
w_{12}&=&w_{1}+w_{2}-w,\\
X_{2}&=&Y+P,\\
Y_{1}&=&X+P,
\\P&=&-\frac{(X-Y)[(p-r)(w-w_{2})+(q-r)(w-w_{1})]}
{(p-r)(w-w_{2})-(q-r)(w-w_{1})-(w-w_{1})(w-w_{2})(X-Y)}.
\end{eqnarray}
\end{subequations}

The parameters $p,q,r$ appearing in the solutions come from the
construction of the models: they are present in the Lax pairs we
used.

The sequence of degrees for the bond variables is
\begin{eqnarray} \label{H1_seq}
\{ d_n\}_b = 1, \; 4 , \; 13 , \; 28 , \; 49 , \; 76 , \; 109 , \; 148
, \; 193 , \; 244 , \; 301 \dots
\end{eqnarray}
The generating function of the  sequence (\ref{H1_seq}) is 
\begin{eqnarray*} 
\zeta(s) = {\frac {1+4\,{s}^{2}+s}{ \left(1- s \right) ^{3}}}
\end{eqnarray*}
The sequence has quadratic growth and the entropy vanishes.

\paragraph{Simplified form:}
By taking suitable linear combinations of  equations \eqref{H1ex-ep0}
we can also write the equations in the form
\begin{subequations}\label{H1x-s}
\begin{eqnarray}
 w_{12}-w_1-w_2 +w &=&0,\\
X + X_2  - Y - Y_1&=&0,\\ \label{H1x_coupling}
X\,X_2- Y\,Y_1- (X-X_2)\frac{q-r}{w-w_2}
+ (Y-Y_1)\frac{p-r}{w-w_1}&=&0.
\end{eqnarray}
\end{subequations}

The equations \eqref{H1x-s} have the form typical for most of our
results: an independent linear or linearizable $w$ equation, a linear
or linearizable equation for $X,Y,X_2,Y_1$, and a coupling
equation. It is then easy to see that the evolution is rational in any
direction: solve the $w$ of any corner from the first equation, then its
adjacent $X,Y$ variables can be solved rationally from the remaining
two.

\paragraph{Remark:} The parameters appearing in the results (here and
below) follow from the Lax set-up \cite{HV11a}, $p,q,r$ corresponding
to the three coordinate directions of the consistency cube. In the end
the regular solution typically depends only on $p-q$, while the exotic
solution depends on $p-r$ and $q-r$. Clearly any finite $r$ can be
absorbed into $p,q$. The limit $r\to\infty$ is also possible, but in
the present case it produces a model that is linear in $X,Y$.

\subsection{H1$\epsilon$}
This quad map was introduced in \cite{ABS-FAA}. There in fact are two kinds
of integrable models of type H1$\epsilon$, related by inversion.
\begin{eqnarray}
(x-x_{12})(x_{1}-x_2)-(p-q)(1+\epsilon x x_{12})=0\label{H1e-der}\\
(x-x_{12})(x_{1}-x_2)-(p-q)(1+\epsilon x_1 x_2)=0,\label{H1e-dss}
\end{eqnarray}
 When extended to the whole lattice they must alternate thereby
 forming a black-white (checkerboard) lattice \cite{XePa09,HV11a}. 

 We may consider two different configurations for the $2\times 2$
 basic cell
\begin{eqnarray*}
%\alpha = 
\setlength{\unitlength}{3947sp}%
\begingroup\makeatletter\ifx\SetFigFont\undefined%
\gdef\SetFigFont#1#2#3#4#5{%
  \reset@font\fontsize{#1}{#2pt}%
  \fontfamily{#3}\fontseries{#4}\fontshape{#5}%
  \selectfont}%
\fi\endgroup%
\begin{picture}(1224,1224)(2089,-1873)
\thinlines
{\color[rgb]{0,0,0}\put(2101,-1261){\framebox(600,600){}}
}%
{\color[rgb]{0,0,0}\put(2701,-1261){\framebox(600,600){}}
}%
{\color[rgb]{0,0,0}\put(2101,-1861){\framebox(600,600){}}
}%
{\color[rgb]{0,0,0}\put(2701,-1861){\framebox(600,600){}}
}%
\put(2280,-1000){\makebox(0,0)[lb]{\smash{{\SetFigFont{12}{14.4}{\rmdefault}{\mddefault}{\updefault}{\color[rgb]{0,0,0}(\ref{H1e-dss})}%
}}}}
\put(1700,-1300){\makebox(0,0)[lb]{\smash{{\SetFigFont{12}{14.4}{\rmdefault}{\mddefault}{\updefault}{\color[rgb]{0,0,0}$\alpha:$}%
}}}}
\put(2880,-1000){\makebox(0,0)[lb]{\smash{{\SetFigFont{12}{14.4}{\rmdefault}{\mddefault}{\updefault}{\color[rgb]{0,0,0}(\ref{H1e-der})}%
}}}}
\put(2280,-1640){\makebox(0,0)[lb]{\smash{{\SetFigFont{12}{14.4}{\rmdefault}{\mddefault}{\updefault}{\color[rgb]{0,0,0}(\ref{H1e-der})}%
}}}}
\put(2880,-1640){\makebox(0,0)[lb]{\smash{{\SetFigFont{12}{14.4}{\rmdefault}{\mddefault}{\updefault}{\color[rgb]{0,0,0}(\ref{H1e-dss})}%
}}}}
\end{picture}%
 \hskip 1truecm   \phantom{and} \hskip 1truecm
%\beta =
\setlength{\unitlength}{3947sp}%
\begingroup\makeatletter\ifx\SetFigFont\undefined%
\gdef\SetFigFont#1#2#3#4#5{%
  \reset@font\fontsize{#1}{#2pt}%
  \fontfamily{#3}\fontseries{#4}\fontshape{#5}%
  \selectfont}%
\fi\endgroup%
\begin{picture}(1224,1224)(2089,-1873)
\thinlines
{\color[rgb]{0,0,0}\put(2101,-1261){\framebox(600,600){}}
}%
{\color[rgb]{0,0,0}\put(2701,-1261){\framebox(600,600){}}
}%
{\color[rgb]{0,0,0}\put(2101,-1861){\framebox(600,600){}}
}%
{\color[rgb]{0,0,0}\put(2701,-1861){\framebox(600,600){}}
}%
\put(1700,-1300){\makebox(0,0)[lb]{\smash{{\SetFigFont{12}{14.4}{\rmdefault}{\mddefault}{\updefault}{\color[rgb]{0,0,0}$\beta:$}%
}}}}
\put(2280,-1000){\makebox(0,0)[lb]{\smash{{\SetFigFont{12}{14.4}{\rmdefault}{\mddefault}{\updefault}{\color[rgb]{0,0,0}(\ref{H1e-der})}%
}}}}
\put(2880,-1000){\makebox(0,0)[lb]{\smash{{\SetFigFont{12}{14.4}{\rmdefault}{\mddefault}{\updefault}{\color[rgb]{0,0,0}(\ref{H1e-dss})}%
}}}}
\put(2280,-1640){\makebox(0,0)[lb]{\smash{{\SetFigFont{12}{14.4}{\rmdefault}{\mddefault}{\updefault}{\color[rgb]{0,0,0}(\ref{H1e-dss})}%
}}}}
\put(2880,-1640){\makebox(0,0)[lb]{\smash{{\SetFigFont{12}{14.4}{\rmdefault}{\mddefault}{\updefault}{\color[rgb]{0,0,0}(\ref{H1e-der})}%
}}}}
\end{picture}%
\end{eqnarray*}

To each of these patterns corresponds a zero curvature condition.
Each of these again have two rational solutions, the regular one and
an exotic one. The exotic solutions are as follows:

\subsubsection{H1$\epsilon\alpha$}
For configuration $\alpha$,  we have:
\paragraph{Original form:}
\begin{subequations}\label{h1alo}
\begin{align}
 X_{2}=&Y+P/Q, \quad Y_1=X+P/Q,\\
P=&(X-Y) [(p-r) (w-w_2) (1+\epsilon  w w_1)+(q-r) (w-w_1)
  (1+\epsilon  w w_2)]\nonumber\\
&+\epsilon  (p+q-2 r) (p-q) (w-w_1) (w-w_2)\\
Q=&
(w-w_1) (w-w_2) (X-Y)\nonumber\\
&-(p-r) (w-w_2) (1+\epsilon  w w_1)+(q-r) (w-w_1) (1+\epsilon  w w_2)\\
 \label{alpha_w}
w_{12}+w=&\frac{(w_1+w_2)(1+\epsilon w^2)} 
{1+\epsilon (w w_1+w w_2-w_1w_2)}.
\end{align}
\end{subequations}

\paragraph{Entropy:}
The sequence of degrees are respectively:
\begin{eqnarray} \label{H1ex_vert_seq}
\{ d_n\}_v = 1 ,\; 3,\; 5,\; 7,\; 9, \; 11,\; 13,\; 15,\; 17,\; 19,\;
21,\dots
\end{eqnarray} 
for the vertex variables, and 
\begin{eqnarray} \label{H1ex_bond_seq}
\{ d_n \}_b = 1,\; 5,\; 17,\; 37,\; 65,\; 101,\; 145,\; 197,\; 257,\;
325, 401,\; 485,\; 577 ,\dots
\end{eqnarray}
for the bond variables.  The sequence (\ref{H1ex_vert_seq}) has linear
growth, signaling the linearizability of the evolution of the vertex
variables (see below). The sequence (\ref{H1ex_bond_seq}) has the
generating function
\begin{eqnarray}
\zeta(s)= \sum_n d_n s^n = {\frac {1+5\,{s}^{2}+2\,s}{ \left(1-s \right) ^{3}}}
\end{eqnarray}
which means quadratic growth of the degrees, i.e. integrability.

\paragraph{Simplified form:} 
Transforming the vertex variables by the Moebius transformation $
w\mapsto (w-1)/[\kappa (w+1)]$ with $\epsilon=-\kappa^2$ etc,
takes (\ref{alpha_w}) into
\begin{equation}\label{H1x1-w}
w \, w_{12}=w_1w_2.
\end{equation}
Then we can write the equations in the form
\begin{subequations}\label{H1x-al}
\begin{eqnarray}
w \, w_{12} -   w_1w_2&=&0,\\
X_2+X-Y_1-Y &=& 0,\\  \label{H1epsilonalpha_coupling}
  X_2X-Y_1Y  - (X-X_2)(q-r)\kappa\frac{w+w_2}{w-w_2}
+(Y-Y_1)(p-r)\kappa\frac{w+w_1}{w-w_1}\nonumber &&\\
+\kappa^2[(p-r)^2-(q-r)^2]&=&0.
\end{eqnarray}
\end{subequations}
In comparison with \eqref{H1x-s}, which was linear in $w$, this is
multiplicative.

\subsubsection{H1$\epsilon\beta$}
For the configuration $\beta$ we have the exotic solution:

\paragraph{Original form:}
\begin{subequations}\label{h1ex1}
\begin{align}
&X_{2}=Y+P_2/Q_2,\quad Y_1=X+P_1/Q_1\\
P_1=&(X-Y)(1+\epsilon X^2)[(p-r)(w-w_2)+(q-r)(w-w_1)],\\
Q_1=&(X-Y) (w-w_1) (w-w_2)-(p-r) (w-w_2) (1+\epsilon  X^2)\nonumber\\
&+(q-r) (w-w_1) (1-\epsilon  X (X-2 Y)),\\
P_2=&(X-Y)(1+\epsilon Y^2)[(p-r)(w-w_2)+(q-r)(w-w_1)],\\
Q_2=&(X-Y) (w-w_1) (w-w_2)
-(p-r) (w-w_2) (1-\epsilon  Y (Y-2 X))\nonumber\\
&+(q-r) (w-w_1) (1+\epsilon  Y^2),\\
w_{12}=&w_1+w_2-w.
\end{align}
\end{subequations}

\paragraph{Entropy:}
The entropy calculation leads to the same conclusion as for the
configuration $\alpha$: the vertex evolution is linear, and
independent of the bonds. The degree sequence for the bonds is the
same as (\ref{H1ex_bond_seq}).

\paragraph{Simplified form:} Starting with \eqref{h1ex1} and using 
the Moebius transformation $X\mapsto  (X-1)/[{\kappa}(X+1)]$,
etc. with $\epsilon=-\kappa^2$, we get:
\begin{subequations}\label{H1x-be}
\begin{eqnarray}
 w_{12}-w_1-w_2+w &=& 0,\\
\label{H1epsilonbeta_coupling}
 X+X_2-Y-Y_1
+2\kappa\frac{q-r}{w-w_2}(X-X_2)
-2\kappa\frac{p-r}{w-w_1}(Y-Y_1)
 &=& 0,\\
 X X_2-Y Y_1 &=& 0. 
\end{eqnarray}
\end{subequations}

\paragraph{Remarks:} H1$\epsilon\alpha$ and H1$\epsilon\beta$ are
deformations that in the limit $\epsilon \to 0$ reduce back to H1. As
a consequence the original forms \eqref{h1alo} and \eqref{h1ex1}
reduce to \eqref{H1ex-ep0}, as can be readily verified. However, the
simplified forms \eqref{H1x-al} and \eqref{H1x-be} are obtained with a
transformation that is singular in $\epsilon$. In these equations
$\epsilon$ appears through $\kappa$ and the $\kappa\to 0$ limit
trivializes them.

The models again depend on $p-r$ and $q-r$, so $r$ can be eliminated,
except that $r\to\infty$ is a possible limit.  In the
H1$\epsilon\beta$ case the $r\to\infty$ limit leads to a nonlinear
model, which may be interesting on its own.

\subsection{dmKdV  ($H3_{\delta=0}$)\label{S3.3}}
If we start from the discrete modified KdV equation ($H3$ with
$\delta=0$ in~\cite{ABS}) we obtain the following:

\paragraph{Original form:}
\begin{subequations}
\begin{eqnarray}
w_{12} w&=&w_1 w_2,\\
X_2 X&=&Y_1 Y,\\
\frac{X_2}{Y}=\frac{Y_1}{X}&=&
\frac{(q^2 w_2-r^2 w)(w-w_1)p X-(p^2 w_1-r^2 w)(w-w_2)q Y}
{(r^2w_2-q^2 w)(w_1-w)p Y-
(r^2w_1-p^2 w)(w_2-w)q X}
\end{eqnarray}
\end{subequations}

\paragraph{Entropy:}
The sequence of degrees we get are respectively
\begin{eqnarray}
\{ d_n\}_v = 1 ,\; 2,\; 3,\; 4,\; 5,\; 6,\; 7, \;8,\; 9,\; 10,\; 11 ,
\dots
\end{eqnarray}
for the vertex variables, and
\begin{eqnarray}
\{d_n\}_b = 1 ,\; 4 ,\; 13 ,\; 28 ,\; 49 ,\; 76 ,\; 109 ,\; 148 ,\;
  193 ,\; 244 ,\; 301, \dots
\end{eqnarray}
for the bond variables, actually the same as for $H1$.
The vertex evolution is linearizable, and the bond evolution is integrable.

\paragraph{Simplified form:}
By taking suitable linear combinations of the original equations we
can write the result also as
\begin{subequations} \label{H3-x}
\begin{eqnarray}
w\,w_{12}-w_1\,w_2&=&0,\\
(X+X_2)p(q^2+r^2)-(X-X_2)p(q^2-r^2)\frac{w+w_2}{w-w_2}\quad &&\nonumber\\
\label{H3x_coupling}
-(Y+Y_1)q(p^2+r^2)+(Y-Y_1)q(p^2-r^2)\frac{w+w_1}{w-w_1}&=&0,\\
X \,X_2-Y\,Y_1&=&0.
\end{eqnarray}
\end{subequations}

\paragraph{Remark:} For H3 the parameter dependence is through $p/r$
and $q/r$. Therefore we may put $r=1$, but the special limits $r=0$ and
$r\to\infty$ also produce interesting sub-cases.

\subsection{dSKdV   ($Q1_{\delta=0}$)  \label{q1exotic}}
This model was obtained from the discrete Schwarzian KdV
equation~\cite{NiQuCa83} ($Q_1$ with $\delta=0$ in~\cite{ABS}).
For the elementary square we used the Lax matrices
\begin{equation}\label{Lax_dskdv}
L(x_1,x;p)=\begin{pmatrix} p x_1 +r(x-x_1) & - p x x_1\\
                         p  &  -p x + r(x-x_1) \end{pmatrix},
%\quad
%M(x_2,x;q)=\begin{pmatrix} q x_2 +r(x-x_2)  &  - q x x_2\\
%                         q & -q x +r ( x -x_2) \end{pmatrix},
\end{equation}
where $L(x_1,x;p)$ describes the parallel transport from $x$ to $x_1$,
and $L(x_2,x;q)$ from $x$ to $x_2$. When constructing the Lax pair for
the bigger $2\times 2$ lattices, we get expressions depending on the
corner variables $x,x_{11},x_{22},x_{1122}$ which will be relabeled as
$w,w_1,w_2,w_{12}$, respectively, while $x_1,x_2,x_{112},x_{122}$ will
become bond variables $X,Y,Y_1,X_2$, respectively.

The results also depend on the three parameters $(p,q,r)$ in an
homogeneous way. The ``true'' parameters are thus $p/r$ and $q/r$.  If
$r$ is not vanishing we do get a rational model, but that model is not
integrable, as we will see. (We will not give here the explicit
solution for $r\neq 0$, as it may be extracted from the simplified form
of the defining equations given at the end of this section.)  In the
$r\rightarrow 0$ limit, we have the following defining relations:
\paragraph{Original form:}
\begin{eqnarray}
  w_{12}& =& \frac{A}{B} \qquad  \mbox {with} \\ \nonumber
  A &=& - \left( { X}-{ Y} \right) \left( 2\,w -X -{ Y} \right) 
{{ w_1}}^{2}{{ w_2}}^{2}+
  \left( -2\,{ Y}\,{{
        X} }^{2}-4\,{ Y}\,{w}^{2}+X{{ Y}}^{2}+4\,w{ Y}\,{ X}+{w}^{ 3}
  \right) {{ w_1}}^{2}{ w_2}
  \\ &&  \nonumber
  -2\,w \left( {
      X}-{ Y} \right) \left( {w}^{2}  -{ Y}\,{ X}\right) { w_1}\,{
    w_2}
  - \left( -2\,{{ Y}}^{2}{ X}-4\,{w}^ {2}{ X}+4\,w{ Y}\,{
      X}+w{{ X}}^{2}+{w}^{3} \right) { w_1}\,{{ w_2}}^{2}
  \\  \nonumber && + {{ Y}}^{2} \left( w-{ X} \right) ^{2}{{
      w_1}}^{2} -{{ X}}^{2}
  \left( w-{ Y} \right) ^{2}{{ w_2}}^{2}
  -w{{ Y}}^{2} \left( w-{ X} \right) ^{2}{ w_1}
  +w{{ X}}^{2} \left( w -{ Y}
  \right) ^{2}{ w_2}, \\ \nonumber
  B& =& - \left( 2\,{w}^{3}{ X}-{{ Y}}^{2}{{
        X}}^{2}+4\,w{ X}\, {{ Y}}^{2}-4\,{w}^{2}{ X}\,{ Y}-{w}^{2}{{
        Y}}^{2} \right) { w_1} -w \left( w-{ X} \right) ^{2}{{
      w_2}}^{2}
  \\ && \nonumber
  + \left( -4\,{w}^{2}{ X}\,{ Y}+2\,{ Y}\,{
      w}^{3}-{{ Y}}^{2}{{ X}}^{2}+4\,w{{ X}}^{2}{ Y}-{{ X }}^{2}{w}^{2}
  \right) { w_2}+w \left( w-{ Y} \right) ^{2}{{ w_1}}^{2}
  \\&&  \nonumber
  +2\, \left( {
      X}-{ Y} \right) \left( {w}^{2} -{ Y}\,{ X}\right) { w_1}\,{ w_2}+
  \left( w-{ X} \right) ^{2 }{ w_1}\,{{ w_2}}^{2}- \left( w-{ Y}
  \right) ^{2}{{ w_1}}^ {2}{ w_2}
  \\ && \nonumber
  +{w}^{2} \left( { X}-{ Y} \right) \left( w{ X}+{ Y}\,w  -2\,{ Y}\,{
      X}\right),
\end{eqnarray}
and 
\begin{eqnarray}
 X_2 &=& \frac{C}{D}  \qquad \mbox{with} \\ \nonumber
C &=& {X}^{2} \left( w-Y \right) ^{2
}{ w_2}
 -{Y}^{2} \left( w-X \right) ^{2}{ w_1}+Y \left( w-X \right) ^{2}{{ w_1}}^{2}
 \\   \nonumber && - \left( 2\,{Y}^{2}X-Y
{X}^{2}-Y{w}^{2}-2\,wYX+2\,{w}^{2}X \right) { w_1}\,{ w_2}+ \left( 
X-Y \right)  \left( -X+2\,w-Y \right) {{ w_1}}^{2}{ w_2},
\\ \nonumber
D &=& - \left( 2\,{Y}^{2}X-Y{X}^{2}-Y{w}^{2}-2\,wYX+2\,{w}^{2}X \right) { 
w_1}
\\  \nonumber  &&
+Y \left( w-X \right) ^{2}{ w_2}+ \left( w-Y \right) ^{2}{{ w_1}
}^{2}- \left( w-X \right) ^{2}{ w_1}\,{ w_2}+w \left( X-Y \right) 
 \left( -2\,YX+wX+wY \right).
\end{eqnarray}
By symmetry, $Y_1$ is obtained from $X_2$ by the exchange $( X \leftrightarrow Y, w_1 \leftrightarrow w_2 )$

\paragraph{Entropy:}
For $r=0$ the sequence of degrees for the vertex variables is 
\begin{equation}
\{ d_n\}_v =  1 , 10, 28, 71, 139, 248, 398, 605, 869, 1206, 1616, 2115, 2703 ,\dots
\end{equation}
fitted by the generating function
\begin{equation}
\zeta_v(s) = {\frac {1 +7\,s +9\,{s}^{3}-{s}^{4}}{ \left(1+ s \right)
    \left(1- s \right) ^{4}}}
\end{equation}

For  the bond variables we get the sequence of degrees
\begin{equation}
\{ d_n\}_b = 1 , 8, 33, 92, 201, 376, 633, 988, 1457, 2056, 2801, 3708 ,
4793, \dots
\end{equation}
fitted by
\begin{equation}
\zeta_b(s) = {\frac { \left( 1+s \right)  \left( 1 +3\,s+ 4\,{s}^{2} \right) }{
 \left( 1-s \right) ^{4}}}
\end{equation}
The growth of the degrees is {\em cubic}, showing integrability. It is
interesting to notice that this example does not have the quadratic
growth so commonly observed.  The vertex and bond variables are
moreover intertwined in a non-trivial way.

The behaviour when $r=0$ has to be contrasted with the generic $r\neq
0$ behaviour, where the degree calculation yields for the vertex
variables
\begin{eqnarray*}
\{ d_n\}_v = 1,\; 10,\; 38,\; 149,\; 565,\; 2110,\; 7882,\; 29425,\;
109817,\; 409850, \; 1529590, \dots
\end{eqnarray*}
and for the bond variables
\begin{eqnarray*}
\{ d_n\}_b = 1, \; 8 , \; 45 , \; 186 , \; 711 , \; 2672 , \; 9991 ,
\; 37304 , \; 139239, \; 519666 , \; 1939437 , \dots
\end{eqnarray*}
fitted respectively by the generating functions
\begin{eqnarray*}
\zeta_v(s) = {\frac {1+6\,s-{s}^{2}+6\,{s}^{3}+{s}^{4}}{ \left( 1-s
    \right) \left(1+s+ {s}^{2} \right) \left(  1 -4\,s+ {s}^{2} \right)
}}, \quad
\zeta_b(s) = {\frac {1+14\,{s}^{2}+13\,{s}^{3}+8\,{s}^{4}+4\,s}{
    \left( 1-s \right) \left( 1+s+{s}^{2} \right) \left(
     1  -4\,s+ {s}^{2} \right) }},
\end{eqnarray*}
 indicating non-integrability (the rate of growth of the degrees
is given by the roots of $s^2-4s+1$, i.e. $2+\sqrt{3}$).

\paragraph{Simplified form:} The transformation
\[
X\mapsto\frac{X w-w_1}{X-1},\,Y\mapsto\frac{Y w-w_2}{Y-1},\,
X_2\mapsto\frac{X_2 w_2-w_{12}}{X_2-1},\,Y_1\mapsto\frac{Y_1 w_1-w_{12}}{Y_1-1},
\]
 simplifies the equations to
\begin{subequations} \label{Q1-rel}
\begin{eqnarray}
%X^2 (w-w_2) (w_2-w_{12})-Y^2 (w-w_1) (w_1-w_{12})&=&0,\\
X_2Y-Y_1X&=&0,\\
Y_1 Y (w - w_1) + w_{12} - w_2&=&0,\\
X_2 X (w - w_2) + w_{12} - w_1&=&0,
\end{eqnarray}
\end{subequations} 
Here is also, for reference, the (nonintegrable) form before
taking the $r\to 0$ limit:
\begin{eqnarray*}
X_2Y-Y_1X&=&r(AX_2+BY_1),\\
Y_1 Y (w - w_1) + w_{12} - w_2&=&0,\\
X_2 X (w - w_2) + w_{12} - w_1&=&0,
\end{eqnarray*}
where
\begin{eqnarray*}
A&=&\frac{(w-w_1)(w-w_2)\, q \, X
  Y-((w-w_2)X-w_1+w_2)^2p}{p\, q\, (w-w_1)(w_1-w_2)},\\
B&=&\frac{(w-w_1)(w-w_2)\, p \, X
  Y - ((w-w_1)Y-w_1+w_2)^2q}{p\, q\, (w-w_2)(w_1-w_2)}.
\end{eqnarray*}

\section{Non-autonomous functional Yang-Baxter equations, 
three dimensional consistency} \label{nonautoYB}

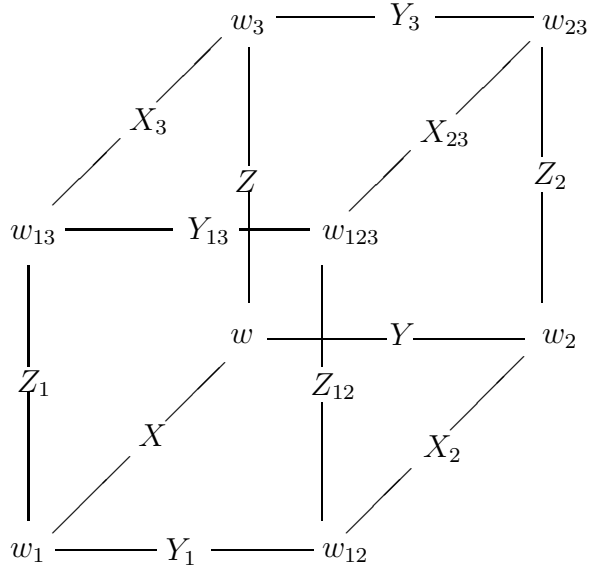
\begin{figure}[h!]
\begin{center}
\setlength{\unitlength}{3000sp}%
\begingroup\makeatletter\ifx\SetFigFont\undefined%
\gdef\SetFigFont#1#2#3#4#5{%
  \reset@font\fontsize{#1}{#2pt}%
  \fontfamily{#3}\fontseries{#4}\fontshape{#5}%
  \selectfont}%
\fi\endgroup%
\begin{picture}(4380,4665)(2536,-6451)
\thinlines
{\color[rgb]{0,0,0}\put(2901,-3461){\line( 1, 1){550}}}%
{\color[rgb]{0,0,0}\put(3751,-2611){\line( 1, 1){530}}
}%
{\color[rgb]{0,0,0}\put(2901,-6161){\line( 1, 1){630}}
}%
{\color[rgb]{0,0,0}\put(3826,-5236){\line( 1, 1){480}}
}%
{\color[rgb]{0,0,0}\put(5301,-6161){\line( 1, 1){530}}
}%
{\color[rgb]{0,0,0}\put(6151,-5311){\line( 1, 1){600}}
}%
{\color[rgb]{0,0,0}\put(4501,-2161){\line( 0,-1){975}}
}%
{\color[rgb]{0,0,0}\put(6901,-2161){\line( 0,-1){900}}
}%
{\color[rgb]{0,0,0}\put(6901,-3361){\line( 0,-1){900}}
}%
{\color[rgb]{0,0,0}\put(4726,-1911){\line( 1, 0){800}}
}%
{\color[rgb]{0,0,0}\put(6026,-1911){\line( 1, 0){800}}
}%
{\color[rgb]{0,0,0}\put(2926,-6301){\line( 1, 0){825}}
}%
{\color[rgb]{0,0,0}\put(4201,-6301){\line( 1, 0){825}}
}%
{\color[rgb]{0,0,0}\put(4651,-4561){\line( 1, 0){975}}
}%
{\color[rgb]{0,0,0}\put(5851,-4561){\line( 1, 0){900}}
}%
{\color[rgb]{0,0,0}\put(2701,-3961){\line( 0,-1){825}}
}%
{\color[rgb]{0,0,0}\put(2701,-5011){\line( 0,-1){1050}}
}%
{\color[rgb]{0,0,0}\put(5101,-3961){\line( 0,-1){825}}
}%
{\color[rgb]{0,0,0}\put(5101,-5086){\line( 0,-1){975}}
}%
{\color[rgb]{0,0,0}\put(5326,-3511){\line( 1, 1){500}}
}%
{\color[rgb]{0,0,0}\put(6201,-2711){\line( 1, 1){600}}
}%
{\color[rgb]{0,0,0}\put(4501,-3361){\line( 0,-1){900}}
}%
{\color[rgb]{0,0,0}\put(3001,-3661){\line( 1, 0){875}}
}%
{\color[rgb]{0,0,0}\put(4426,-3661){\line( 1, 0){575}}
}%
\put(6901,-4600){\makebox(0,0)[lb]{\smash{{\SetFigFont{12}{14.4}%
{\rmdefault}{\mddefault}{\updefault}{\color[rgb]{0,0,0}$w_2$}%
}}}}
\put(4351,-4600){\makebox(0,0)[lb]{\smash{{\SetFigFont{12}{14.4}%
{\rmdefault}{\mddefault}{\updefault}{\color[rgb]{0,0,0}$w$}%
}}}}
\put(2551,-6361){\makebox(0,0)[lb]{\smash{{\SetFigFont{12}{14.4}%
{\rmdefault}{\mddefault}{\updefault}{\color[rgb]{0,0,0}$w_1$}%
}}}}
\put(5101,-6361){\makebox(0,0)[lb]{\smash{{\SetFigFont{12}{14.4}%
{\rmdefault}{\mddefault}{\updefault}{\color[rgb]{0,0,0}$w_{12}$}%
}}}}
\put(5656,-1976){\makebox(0,0)[lb]{\smash{{\SetFigFont{12}{14.4}%
{\rmdefault}{\mddefault}{\updefault}{\color[rgb]{0,0,0}$Y_3$}%
}}}}
\put(3526,-2836){\makebox(0,0)[lb]{\smash{{\SetFigFont{12}{14.4}%
{\rmdefault}{\mddefault}{\updefault}{\color[rgb]{0,0,0}$X_3$}%
}}}}
\put(6826,-3286){\makebox(0,0)[lb]{\smash{{\SetFigFont{12}{14.4}%
{\rmdefault}{\mddefault}{\updefault}{\color[rgb]{0,0,0}$Z_2$}%
}}}}
\put(5656,-4616){\makebox(0,0)[lb]{\smash{{\SetFigFont{12}{14.4}%
{\rmdefault}{\mddefault}{\updefault}{\color[rgb]{0,0,0}$Y$}%
}}}}
\put(5926,-5536){\makebox(0,0)[lb]{\smash{{\SetFigFont{12}{14.4}%
{\rmdefault}{\mddefault}{\updefault}{\color[rgb]{0,0,0}$X_2$}%
}}}}
\put(3826,-6386){\makebox(0,0)[lb]{\smash{{\SetFigFont{12}{14.4}%
{\rmdefault}{\mddefault}{\updefault}{\color[rgb]{0,0,0}$Y_1$}%
}}}}
\put(3601,-5446){\makebox(0,0)[lb]{\smash{{\SetFigFont{12}{14.4}%
{\rmdefault}{\mddefault}{\updefault}{\color[rgb]{0,0,0}$X$}%
}}}}
\put(2606,-4996){\makebox(0,0)[lb]{\smash{{\SetFigFont{12}{14.4}%
{\rmdefault}{\mddefault}{\updefault}{\color[rgb]{0,0,0}$Z_1$}%
}}}}
\put(4386,-3356){\makebox(0,0)[lb]{\smash{{\SetFigFont{12}{14.4}%
{\rmdefault}{\mddefault}{\updefault}{\color[rgb]{0,0,0}$Z$}%
}}}}
\put(5006,-5021){\makebox(0,0)[lb]{\smash{{\SetFigFont{12}{14.4}%
{\rmdefault}{\mddefault}{\updefault}{\color[rgb]{0,0,0}$Z_{12}$}%
}}}}
\put(5901,-2936){\makebox(0,0)[lb]{\smash{{\SetFigFont{12}{14.4}%
{\rmdefault}{\mddefault}{\updefault}{\color[rgb]{0,0,0}$X_{23}$}%
}}}}
\put(4001,-3736){\makebox(0,0)[lb]{\smash{{\SetFigFont{12}{14.4}%
{\rmdefault}{\mddefault}{\updefault}{\color[rgb]{0,0,0}$Y_{13}$}%
}}}}
\put(5101,-3736){\makebox(0,0)[lb]{\smash{{\SetFigFont{12}{14.4}%
{\rmdefault}{\mddefault}{\updefault}{\color[rgb]{0,0,0}$w_{123}$}%
}}}}
\put(6901,-2011){\makebox(0,0)[lb]{\smash{{\SetFigFont{12}{14.4}%
{\rmdefault}{\mddefault}{\updefault}{\color[rgb]{0,0,0}$w_{23}$}%
}}}}
\put(4351,-2011){\makebox(0,0)[lb]{\smash{{\SetFigFont{12}{14.4}%
{\rmdefault}{\mddefault}{\updefault}{\color[rgb]{0,0,0}$w_3$}%
}}}}
\put(2551,-3736){\makebox(0,0)[lb]{\smash{{\SetFigFont{12}{14.4}%
{\rmdefault}{\mddefault}{\updefault}{\color[rgb]{0,0,0}$w_{13}$}%
}}}}
\end{picture}%
\end{center}
 \caption{The consistency cube \label{FigYB}}
\end{figure}

One interesting property of the four models proposed in
Sec.~\ref{S3.1}-\ref{S3.3} is that they provide non-autonomous
solutions to the functional form of the Yang-Baxter
equations~\cite{Dr92}, alias Yang-Baxter maps. This may be seen by
examining the 3 dimensional consistency of the models on the
configuration shown in Figure \ref{FigYB}. The vertices introduce
position dependence, and the maps induced on the bonds verify a
modified functional Yang-Baxter equations if the model is 3D
consistent.

Suppose that $w,w_1,w_2,w_3,X,Y,Z$ are given. Then the relation on
each face allows one to evaluate $X_2,X_3,Y_1,Y_2,Z_1,Z_2 ,w_{12}
,w_{13} ,w_{23}$ in a unique way, $X_{23},Y_{13},Z_{12}$ in two
different ways, and $w_{123}$ may be computed in three different ways.
Consistency means they all have to give the same results.{\em We have
  checked that all the integrable models described here verify this 3
  dimensional consistency.}

It should be noted that for the model presented in section
\ref{q1exotic}, the consistency is not verified when $r\neq 0$,
precisely in the case where it is not integrable, but is verified for
$r=0$ (that is to say the limit $p=q=s=\infty$), where it is
integrable, as ascertained by the vanishing of the algebraic entropy.

The first four sets of simplified equations have similar
structure. The vertex equation is independent of the bonds, and the
values $w$ might be considered as parameters. They introduce
non-autonomy in the bond evolution, which turns out to be the
functional equivalent of the quantum Yang-Baxter equation introduced
in~\cite{GeNe1984}.  The vertex equations are either linear (additive
case), or linearizable (multiplicative case), and can be solved
explicitly.  In the additive case $w +w_{12} - w_1 -w_2 =0$ the
general solution is $ w_{n,m} = F(n) + G(m)$, and in the
multiplicative case $w \, w_{12} - w_1 \, w_2 =0$, we have $ w_{n,m} =
F(n) \, G(m)$, with $F$ and $G$ arbitrary functions.  The equations
determining $X$ and $Y$ contain one linear part, and a coupling
equation which introduces non-linearity.

\subsection{dpKdV and $H1\epsilon$, configuration $\alpha$ }

We can solve dpKdV \eqref{H1x-s} and $H1\epsilon$, configuration
$\alpha$, \eqref{H1x-al} together.  The vertex evolution is
additive for (\ref{H1x-s}a) and multiplicative for (\ref{H1x-al}a) and
is solved as described above.  Equations (\ref{H1x_coupling}) and
(\ref{H1epsilonalpha_coupling}) can then be cast in the common form:
\begin{eqnarray} \label{XYmult}
X \, X_2 - Y Y_1  -  g(m)\;  (X - X_2) + f(n)\;  (Y -Y_1) + \omega =0,
\end{eqnarray} 
where $f$ and $g$ are readily expressible in terms of the
aforementioned $F(n)$ and $G(m)$ and $\omega$ is a constant.

We can solve \eqref{XYmult} together with the constraint $ X + X_2 = Y
+ Y_1$ for arbitrary $f(n),g(m),\omega$, obtaining
\begin{eqnarray*}
  X_2  = Y + R, \qquad Y_1  =  X + R, \qquad  \displaystyle
  \mbox{with } \qquad  \displaystyle R  =
  \frac{f(n)^2-g(m)^2-\omega}{Y-X+f(n)-g(m)}.  
\end{eqnarray*}

Redefining the $X$ and $Y$ by the shifts
\begin{eqnarray*}
X  = X' + f(n) + \sigma(n)+\rho(m), \qquad
Y  = Y' + g(m) + \sigma(n)+\rho(m),
\end{eqnarray*} 
where $\sigma,\rho$ are determined from $\sigma(n+1)-\sigma(n) = 2 \,
f(n)$ and $\rho(m+1)-\rho(m) = 2 \, g(m)$, we finally get
\begin{eqnarray}
 X'_2  = Y' + P, \qquad Y'_1  =  X' + P,  \qquad
\mbox{with } \qquad \displaystyle   P  =  \frac{f(n)^2-g(m)^2-\omega}{X'-Y'}.
\end{eqnarray}
This is a non-autonomous version of the Adler map \cite{AdMap} ($F_V$
in the classification of~\cite{ABS-CAG}).

\subsection{ $H1\epsilon$, configuration $\beta$ and dmKdV ($H3_{\delta=0}$)}
Equations \eqref{H1x-be} and \eqref{H3-x} can also be solved together.
In both cases the coupling equations (\ref{H1epsilonbeta_coupling}) and
(\ref{H3x_coupling}) can be written as
\begin{eqnarray} \label{XYadd}
\mu \; ( X + X_2) - \nu\; ( Y + Y_1 ) - g(m)\; (X - X_2) + f(n)\; (Y
-Y_1) =0,
\end{eqnarray} 
where $f$ and $g$ are, as above, some functions related to
$F(n),G(m)$, and $\mu$ and $\nu$ are constants.  Equation
\eqref{XYadd} together with the constraint $X \; X_2 - Y \; Y_1=0$ is
solved in the generic case by
\begin{equation}
X_2 = Y \; Q, \qquad Y_1 = X \; Q, \qquad Q = \frac{(\mu-g(m)) \, X -  (
 \nu - f(n)) \, Y }{ (\nu + f(n)) \, X -  (\mu +g(m)) \; Y},
\end{equation}

Scaling $X$ and $Y$ by 
\begin{equation}
X=X'\sigma(n)\rho(m)/(f(n)+\nu),\qquad
Y=Y'\sigma(n)\rho(m)/(g(m)+\mu),
\end{equation}
where $\sigma,\rho$  now solve 
\begin{equation}
\sigma(n+1)=\sigma(n)\;\frac{\nu-f(n)}{\nu+f(n)} \qquad
\rho(m+1)=\rho(m)\;\frac{\mu-g(m)}{\mu+g(m)},
\end{equation}
we get the equations in the form
\begin{eqnarray}\label{YB-F3}
X'_2=\frac{Y'}{\alpha(n)}P,\quad 
Y'_1=\frac{X'}{\beta(m)}P,\quad
P=\frac{\alpha(n) X'-\beta(m)Y'}
{X'-Y'},
\end{eqnarray}
with
\begin{eqnarray*}
  \alpha(n) = \frac{ 1 }{ \nu^2 - f(n)^2 }, \qquad \beta(m) = \frac{ 1  }{ \mu^2 - g(m)^2}.  
\end{eqnarray*}
Equation \eqref{YB-F3} is a non-autonomous version of $F_{III}$ of
\cite{ABS-CAG}.

It should be noticed that the non-autonomous nature of the above
equations parallels the results of~\cite{SaRaHy07}.

\subsection{dSKdV   ($Q1_{\delta=0}$)}
In the case of equation \eqref{Q1-rel} the $w$ evolution is not
linearizable, as indicated by the algebraic entropy analysis. If we
consider  \eqref{Q1-rel} as three equations for two variables $X,Y$
then we can derive from their compatibility an equation for $w$, which
can be written as
\begin{subequations}\label{Q1w}
\begin{equation}
\frac{T(n,m+2)}{T(n,m)}=\frac{H(n+2,m)}{H(n,m)},
\end{equation}
where $T,H$ are 3-point Schwarzian-like derivatives of $n,m$, respectively:
\begin{equation}\label{q1weq}
T(n,m):=\frac{w(n+2,m)-w(n+1,m)}{w(n+1,m)-w(n,m)},\quad
H(n,m):=\frac{w(n,m+2)-w(n,m+1)}{w(n,m+1)-w(n,m)}.
\end{equation}
\end{subequations}
Equation \eqref{Q1w} connects the points of a $3\times 3$ sublattice,
except the center point (in the above $w(n+1,m+1)$). The algebraic
entropy analysis indicates the rare cubic growth. Equations defined
on a $3\times 3$ sublattice are typical for Boussinesq-type lattice
equations \cite{Bous}, but they usually have quadratic growth.

The Schwarzian KdV equation is given by
\begin{equation}
S(n,m):=\frac{[w(n,m)-w(n+1,m)][w(n,m+1)-w(n+1,m+1)]}{
[w(n,m)-w(n,m+1)][w(n+1,m)-w(n+1,m+1)]}=\frac{p}q.
\end{equation}
From \eqref{Q1w} one can now derive
\begin{equation}
S(n+1,m+1)S(n,m)=S(n+1,m)S(n,m+1),
\end{equation}
which can be solved and we find that $w$ solves the non-autonomous
SKdV equation
\begin{equation}
\frac{[w(n,m)-w(n+1,m)][w(n,m+1)-w(n+1,m+1)]}{
[w(n,m)-w(n,m+1)][w(n+1,m)-w(n+1,m+1)]}=\frac{f(n)}{g(m)}.
\end{equation}

Thus when the model \eqref{Q1-rel} is considered on larger cells made of
four adjacent squares, we find an independent non-trivial evolution
of the vertex variables \eqref{Q1w}, driving the evolution of the bond
variables according to (\ref{Q1-rel}b,c).

\section{Conclusion and perspectives}

We have presented lattice models with values given at both the
vertices and the bonds of a 2D square lattice. They were constructed
using the ambiguity of the zero curvature condition on a coarse
grained lattice.  The integrability of these models is guaranteed by
the vanishing of the algebraic entropy and by the
Consistency-Around-the-Cube property.

Four out of five models provide non autonomous generalizations of
known Yang-Baxter maps, the fifth case \eqref{Q1-rel} (related to
dSKdV) being seemingly different.

They all share some remarkable algebraic properties, in particular
their multi-rationality, which might be the frame for further
examples and an eventual classification.

Many more aspects will have to be examined (especially for
\eqref{Q1-rel}), like the existence of B\"acklund transforms,
proper Lax pairs, symmetries, reductions (periodic reductions as well
as similarity reductions) which should produce interesting integrable
maps, including discrete Painlev\'e equations. Their continuous limits
should also be considered. All these go beyond the scope of this
paper, and require further studies.

\subsection*{Acknowledgments}
One of us (JH) was partially supported by the Ville de  Paris in the
``Research in Paris'' program. Some of the computations were done
using REDUCE \cite{Red}. We would like to thank F. Nijhoff and J. Perk
for additional references.


\begin{thebibliography}{99}

\bibitem{ABS} V Adler, A Bobenko and Yu Suris, {\em Classification of
  Integrable Equations on Quad-Graphs.  The Consistency Approach},
  Commun. Math.  Phys. {\bf 233} (2003) 513--543, {\tt arXiv:nlin/0202024}.

\bibitem{Dr92} V.G. Drinfeld, {\em On some unsolved problems in
  quantum group theory}, in: Quantum Groups, in: Lecture Notes in
  Mathematics, Vol. 1510, Springer, New York, 1992, p. 1.

\bibitem{Ve03}  A.P. Veselov, {\em Yang-Baxter maps and integrable dynamics},
Phys. Lett A {\bf 314} (2003) 214--221.

\bibitem{ABS-CAG} V.E. Adler, A.I. Bobenko and Yu.B. Suris, {\em
  Geometry of Yang–Baxter Maps: pencils of conics and quadrirational
  mappings}, Commun. Anal. Geom., {\bf 12}, 967--1007 (2004), {\tt
  arXiv:math/0307009}.

\bibitem{PaTo07} V.G. Papageorgiou, A.G. Tongas, {\em Yang-Baxter maps
  and multi-field integrable lattice equations}, J. Phys. A:
  Math. Theor. {\bf 40} (2007) 12677--12690, {\tt arXiv:math/0702577}.

\bibitem{JH-B} J. Hietarinta, {\em Boussinesq-like multi-component
  lattice equations and multi-dimensional consistency}, J. Phys. A:
  Math. Theor. {\bf 44} (2011) 165204 (22pp), {\tt arXiv:1011.1978}
% doi:10.1088/1751-8113/44/16/165204

\bibitem{SpNiKa10} P.E.  Spicer, F.W. Nijhoff and P. H. van der Kamp,
  {\em Higher Analogues of the Discrete-Time Toda Equation and the
    Quotient-Difference Algorithm},  {\tt arXiv:1005.0482} , to appear in
  Nonlinearity.

\bibitem{Ni11} F.W.  Nijhoff, {\em A higher-rank version of the $Q3$
  equation}, {\tt arXiv:1104.1166}.

\bibitem{JKMO} M. Jimbo, A. Kuniba, T. Miwa and M. Okado,
  {\em The $A_n^{(1)}$ Face Models},
  Commun. Math. Phys. {\bf 119} (1988) 543--565.

\bibitem{HV11a} J. Hietarinta and C. Viallet, {\em Weak Lax
  pairs for lattice equations} (2011), {\tt arXiv:1105.3329}.

\bibitem{FLax}
FW Nijhoff, {\em Lax pair for the Adler (lattice Krichever-Novikov) system},
Phys. Lett. A {\bf 297} (2002) 49--58,  {\tt arXiv:nlin/0110027}. 

\bibitem{BSLax} A.Bobenko, Yu. Suris, {\em Integrable systems on
  quad-graphs}, IMRN 11 (2002) 573--611, {\tt arXiv:nlin/0110004v1}.

\bibitem{QCRdual} G.R.W. Quispel, H.W. Capel and J.A.G. Roberts. {\em
Duality for discrete integrable systems}, J. Phys. A: Math. Gen. {\bf
38} (2005) 3965--3980.

\bibitem{AE1} M. Bellon and C-M. Viallet. {\em Algebraic Entropy},
  {Comm. Math. Phys.} {\bf 204} (1999) {425--437}, 
{\tt arXiv:chao-dyn/9805006}.

\bibitem{AE2} C-M. Viallet {\em Algebraic entropy for lattice
  equations}, {\tt arXiv:math-ph/0609043}

\bibitem{ABS-FAA} V. E. Adler, A. I. Bobenko, and Yu. B. Suris.  {\em
  Discrete Nonlinear Hyperbolic Equations. Classification of Integrable
  Cases}, Funct. Anal. App., {\bf 43}, 3--17 (2009), {\tt arXiv:0705.1663}.

\bibitem{AdMap} V.E. Adler, {\em Recuttings of polygons},
  Funct. Anal. App., {\bf 27}  (1993) 141--143.

\bibitem{XePa09} P.D. Xenitidis and V.G. Papageorgiou, {\em Symmetries
  and integrability of discrete equations defined on a black–-white
  lattice} J. Phys. A: Math. Theor. {\bf 42} (2009), 454025, {\tt arXiv:0903.3152}.

\bibitem{NiQuCa83} F.W. Nijhoff, G.R.W.  Quispel and H.W. Capel {\em
  Direct linearization of nonlinear difference-difference equations},
  Phys. Lett. A {\bf 97} (1983) 125--128.

\bibitem{GeNe1984} J.-L. Gervais and A. Neveu, {\em Novel Triangle Relation
  and Absence of Tachyons in Liouville String Field Theory},
  Nucl. Phys. {\bf B238} (1984) 125-141.

\bibitem{SaRaHy07} R. Sahadevan, O.G. Rasin, P.E. Hydon {\em
    Integrability conditions for nonautonomous quad-graph equations},
  J. Math. Anal. Appl. {\bf 331} (2007) 712--726.

\bibitem{Bous} F.W. Nijhoff, Discrete Painlev\'e Equations and
  Symmetry Reduction on the Lattice, in: {\em Discrete Integrable
    Geometry and Physics}, eds. A.I. Bobenko and R. Seiler, (Clarendon
  Press, Oxford, 1999), pp 209--234.

\bibitem{Red} A. Hearn, {\em REDUCE User’s Manual Version
  3.8} (2004)  \\ {\tt http://reduce-algebra.sourceforge.net/}.

\end{thebibliography}
\end{document}